\documentclass[12pt]{article}
\usepackage{amsmath}
\usepackage{graphicx}
\usepackage{natbib}
\usepackage{url} 

\newcommand{\blind}{0}

\begin{document}

\def\spacingset#1{\renewcommand{\baselinestretch}%
{#1}\small\normalsize} \spacingset{1}


\if0\blind
{
  \title{\bf A Comprehensive Comparison of the Wald, Wilson, and adjusted Wilson  Confidence Intervals for Proportions}
  \author{Nabil Kahouadji\thanks{
    The author gratefully acknowledges the National Science Foundation (NSF DMS-2331502) for funding this research.}\hspace{.2cm}\\
    Department of Mathematics, Northeastern Illinois University}
  \maketitle
} \fi

\if1\blind
{
  \bigskip
  \bigskip
  \bigskip
  \begin{center}
    {\LARGE\bf A Comprehensive Comparison of the Wald, Wilson, and adjusted Wilson  Confidence Intervals for Proportions}
\end{center}
  \medskip
} \fi

\bigskip
\begin{abstract}

The standard confidence interval for a population proportion  covered in the overwhelming  majority of introductory and intermediate statistics textbooks surprisingly remains the Wald confidence interval despite having a poor coverage probability,  especially for small sample sizes or when the unknown population proportion is close to either 0 or 1. Using the mean coverage probability, and for some sample sizes, Agresti and Coull showed not only that the 95\% Wilson confidence interval performs better, but also showed that 95\% adjusted Wilson of type 4 confidence interval, obtained by simply adding four pseudo-observations, outperforms both the Wald and the Wilson confidence intervals. In this paper, we introduce a rainbow color code and pixel-color plots as ways to comprehensively compare the Wald, Wilson, and adjusted-Wilson of type $\epsilon$ confidence intervals across all sample sizes $n=1, 2, \dots, 1000$, population proportion values $p=0.01, 0.02, \dots, 0.99$, and for the three typical confidence levels. We show not only that adding 3 (resp., 4 and 6) pseudo-observations is the best for the 90\% (resp., 95\% and 99\%) adjusted Wilson confidence interval, but it also performs better than both the 90\% (resp., 95\% and 99\%) Wald and Wilson confidence intervals.

\end{abstract}

\noindent%

\spacingset{1.75} 
\section{Introduction}
\label{sec:intro}

A basic and important statistical inference is the confidence interval for a dichotomous categorical variable. Let $X$ be a binomial random variable with $n$ trials and probability of (a single) success $p$. This probability is a population proportion for which the sample proportion $\hat{p} = X/n$ is a point estimate. The overwhelming majority of statistics and probability textbooks introduce the confidence interval for a population proportion based on the asymptotic normality of the sample proportion and estimating the standard error.  This interval is the $100(1-\alpha)\%$ Wald confidence interval for $p$, and  is defined as
\begin{equation}\label{WaldCI}
\hat{p} \pm z_{\alpha/2}\sqrt{\dfrac{\hat{p}(1-\hat{p})}{n}},
\end{equation}
where $z_{\alpha/2}$ is the $100(1-\alpha/2)$ quantile of the standard normal distribution. This confidence interval results from inverting the Wald test for $p$, that is,  the interval is the set of $p_0$ values having p-value exceeding $\alpha$ in testing $H_0\colon p = p_0$ against $H_a\colon p \neq p_0$ using the test statistic $z = (\hat{p} - p_0)/\sqrt{\hat{p}(1-\hat{p})/n}$. 
The Wald confidence interval, while simple and widely taught and used, has been shown to perform poorly unless  the sample size $n$ is large  (see \cite{Agresti1} and  \cite{Agresti2} for a comprehensive literature review).  A less known, taught and used confidence interval is the $100(1-\alpha)\%$ Wilson confidence interval for $p$, defined as
\begin{equation}\label{WilsonCI}
\dfrac{n\hat{p} + \frac12 z^2_{\alpha/2}}{n + z^2_{\alpha/2} } \pm z_{\alpha/2} \sqrt{ \dfrac{n\hat{p}(1-\hat{p}) + \frac14 z^2_{\alpha/2} }{(n+z^2_{\alpha/2})^2}}
\end{equation}

This confidence interval is due to \cite{Wilson} who was inspired by the law of succession of \cite{Laplace}, and showed that a much better confidence interval for a single proportion is based on inverting the test with standard error evaluated at the null hypothesis, that is the set of $p_0$ values for which $|\hat{p} - p_0|/\sqrt{p_0(1-p_0)/n} < z_{\alpha/2}$.  While the Wilson confidence interval (\ref{WilsonCI}) may seem surprising at first encounter, rewriting it as
\begin{equation}\label{WilsonbisCI}
\bigg(\dfrac{n}{n+ z^2_{\alpha/2}}\bigg)\hat{p}  + \bigg(\frac{z^2_{\alpha/2}}{n+z^2_{\alpha/2}}\bigg) \frac{1}{2} \pm z_{\alpha/2}\sqrt{\dfrac{\bigg(\dfrac{n}{n+ z^2_{\alpha/2}} \bigg) \hat{p}(1-\hat{p}) + \bigg(\dfrac{z^2_{\alpha/2}}{n+z^2_{\alpha/2}} \bigg)  \frac12 (1-\frac12) }{n+ z^2_{\alpha/2}}}
\end{equation}
shows that the midpoint is a weighted average of $\hat{p}$ and $1/2$, and it is equal to the sample proportion after adding $z^2_{\alpha/2}$ pseudo-observations, half of which are successes and the other half are failures. Moreover, the same weights $n/(n+z^2_{\alpha/2})$ and $z^2_{\alpha/2}/(n+z^2_{\alpha/2})$ appear in the square coefficient of $z_{\alpha/2}$, i.e., the Wilson standard error. Indeed, the square of the Wilson standard error is the (same) weighted average of the variance of a sample proportion when $p = \hat{p}$ and the variance of a sample proportion when $p=1/2$ together with a sample size $n+z_{\alpha/2}$ instead of the usual sample size $n$.  For small sample sizes, the Wilson confidence interval is biased toward $p=1/2$. As the sample size increases, the Wilson confidence interval converges to the Wald confidence interval. \cite{Agresti1} not only showed that the performance of the 95\%-Wilson confidence interval, which they called \textit{score confidence interval}, is much better than 95\%-Wald confidence interval, but they introduced also a simple variation of the 95\%-Wald confidence interval by adding two successes and two failures, and then use a standard error with this augmented sample size ($n+4)$ rather than using the Wilson standard error. The augmented sample size by four corresponds simply to the rounding of $z_{0.025}^2 = 1.96^2 \approx 4$. One can consider different values of pseudo-observations to be added to the sample size, thus leading to different adjusted confidence intervals. In order to determine the optimal number of pseudo-observations for various confidence levels, let us define the $100(1-\alpha)\%$ adjusted Wilson of type $\epsilon$, or simply the adjusted Wilson $\epsilon$, confidence interval for $p$ as:  \begin{equation}
\tilde{p} \pm z_{\alpha/2}\sqrt{\dfrac{\tilde{p}(1-\tilde{p})}{n + \epsilon}} \quad \text{ where } \tilde{p} = \dfrac{n\hat{p} + \frac12 \epsilon}{n + \epsilon}
\end{equation}
In other words, the adjusted Wilson of type $\epsilon$ confidence interval simply consists of adding $\epsilon$ pseudo-observations to the sample, half of which are successes, the other half are failures, and then compute the Wald confidence interval with this adjusted point estimate of the population proportion.  In \cite{Agresti1}, the mean coverage probability, which is the average performance across all $p \in (0,1)$, for the 95\% confidence interval of the Wald, Wilson (called Exact) and adjusted Wilson (called adjusted Wald) has been compared for sizes $n=5, 15, 30, 50$ and $100$. While the reported mean coverage probabilities are high, the poor performance of the confidence intervals when $p$ is close to 0 or 1 is hidden as an artifact of the averaging across all $p \in (0,1)$. In this paper, we compare the performance of the Wald, Wilson and adjusted Wilson $\epsilon =1, 2, \dots, 8$ confidence intervals for each one of the three confidence levels $c=90\%, 95\%$ and $99\%$, and for each pair $(n,p)$ where the sample size $n=1, 2, \dots, 100$ (and $n=1, 2, \dots, 1000$) and population proportion $p = 0.01, 0.02, \dots, 0.99$. We also introduce pixel-color plots based on a rainbow color code as a way to comprehensively compare these confidence intervals across all sample sizes and all population proportion values, and to also determine the optimal number of pseudo-observations $\epsilon$ for a given confidence level $c$.




\section{Comparing Confidence Intervals for Proportions}
\label{sec:}

The actual coverage probability of an interval estimator is the probability that the interval estimator contains the population parameter value. Thus,  a probabilistic interpretation of the coverage probability of an interval estimator is the average performance of such interval. Indeed, consider the following experiment: extract a random sample of size $n$ from a fixed population, compute the $100(1-\alpha)\%$ confidence interval for the population parameter, where $\alpha$ is a fixed significance level, and check whether this confidence interval includes the population parameter. If one repeats this experiment a large number of times $N_\text{Sim}$, then the frequency  of the number of times the population parameter belonged to the confidence interval divided by $N_\text{Sim}$ is the (empirical) coverage probability of this interval estimator. This coverage probability should be at least $100(1-\alpha)\%$, otherwise, the method/formula used to compute the $100(1-\alpha)\%$ confidence interval for the chosen sample size and/or method is flawed. 

In order to comprehensively compare  the coverage probabilities of the Wald, Wilson and adjusted Wilson of type $\epsilon$ confidence intervals across all pairs $(n, p)$, with sample sizes $n=1, 2, \dots $ and population proportions $p = 0.01, 0.02, \dots, 0.99$, and to also determine the optimal number of extra pseudo-observations $\epsilon$ for the adjusted Wilson method for a given confidence level, we propose a graphical representation based on the following eight rainbow color code. Given a (fixed) significance level $\alpha$, corresponding to a (fixed) confidence level $c = 1 - \alpha$, the pair $(n,p)$, also called pixel, is colored based on its coverage probability (across $N_\text{Sim}$ simulations) as in Table~\ref{tab:ColorCode}.

\begin{table}[h]
\caption{Coverage probability color code for a significant level $\alpha$  \label{tab:ColorCode}}
\begin{center}
\begin{tabular}{rrrr}
\hline
Coverage Probability& Color & Coverage Probability& Color \\\hline
$[1-\alpha, 1]$ & Pink &$[3\alpha, 0.5)$ & Green \\
$[1-2\alpha, 1-\alpha)$ & Purple & $[2\alpha, 3\alpha)$ & Light Green \\
$[1-3\alpha, 1-2\alpha)$ & Blue & $[\alpha, 2\alpha)$ & Orange\\\
$[0.5, 1-3\alpha)$ & Turquoise & $[0, \alpha)$ & Red \\  \hline
\end{tabular}
\end{center}
\end{table}

\begin{figure}[h]
\begin{center}
\includegraphics[scale=.4]{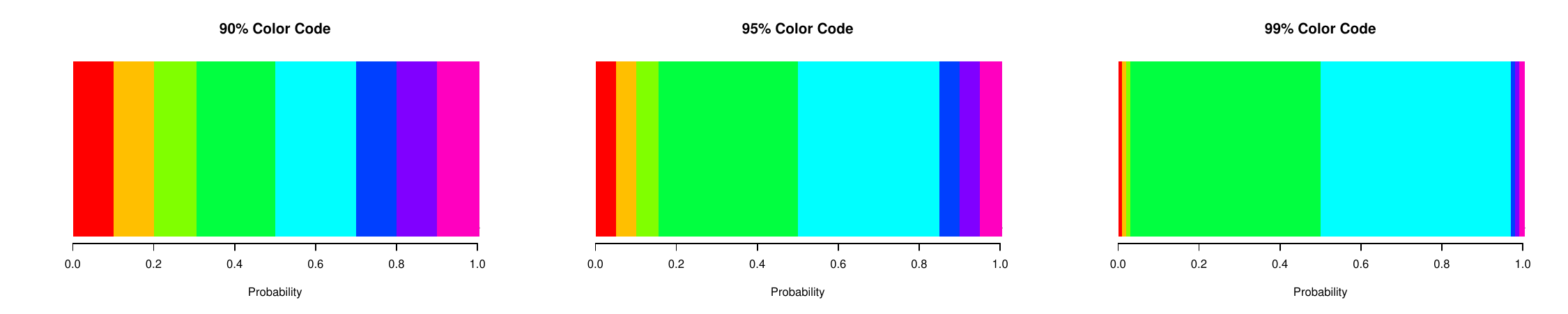}
\end{center}
\caption{Coverage probability color code for the 90\%, 95\% and 99\% confidence Intervals. \label{fig:ColorCode}}
\end{figure}

For typical confidence levels $c = 90\%, 95\%$ and $99\%$, corresponding to significant levels $\alpha = 10\%, 5\%$ and $1\%$ respectively, the rainbow color code in Table \ref{tab:ColorCode}. is shown in Figure~\ref{fig:ColorCode}.

\begin{figure}[h]
\begin{center}
\includegraphics[scale=.42]{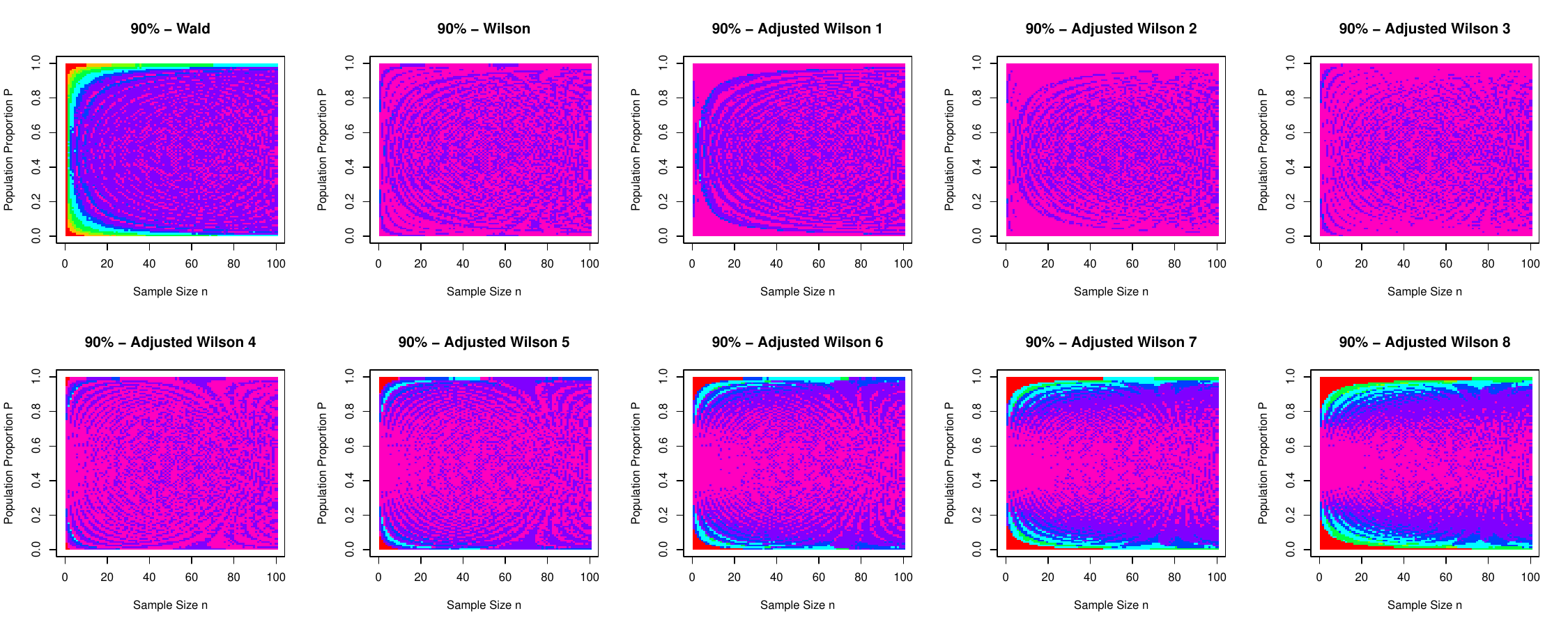}
\end{center}
\caption{Comparison of the 90\% Wald, Wilson and adjusted Wilson confidence intervals' coverage probabilities for pairs $(n,p)$  where $n = 1, 2, \dots, 100$, $p = 0.01, 0.02, \dots, 0.99$}
\label{fig:90comp100}
\end{figure}

\begin{figure}[h]
\begin{center}
\includegraphics[scale=.42]{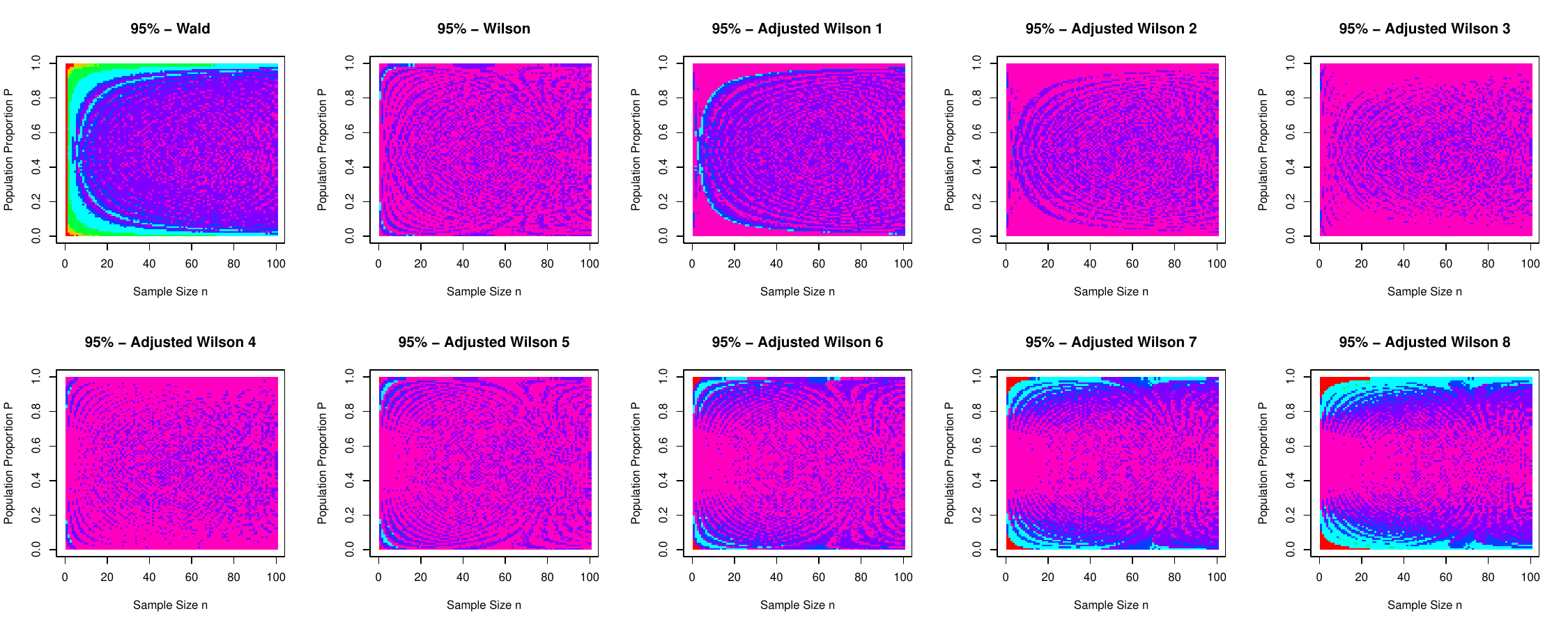}
\end{center}
\caption{Comparison of the 95\% Wald, Wilson and adjusted Wilson confidence interval's coverage probabilities for pairs $(n,p)$  where $n = 1, 2, \dots, 100$, $p = 0.01, 0.02, \dots, 0.99$}
\label{fig:95comp100}
\end{figure}

\begin{figure}[h]
\begin{center}
\includegraphics[scale=.42]{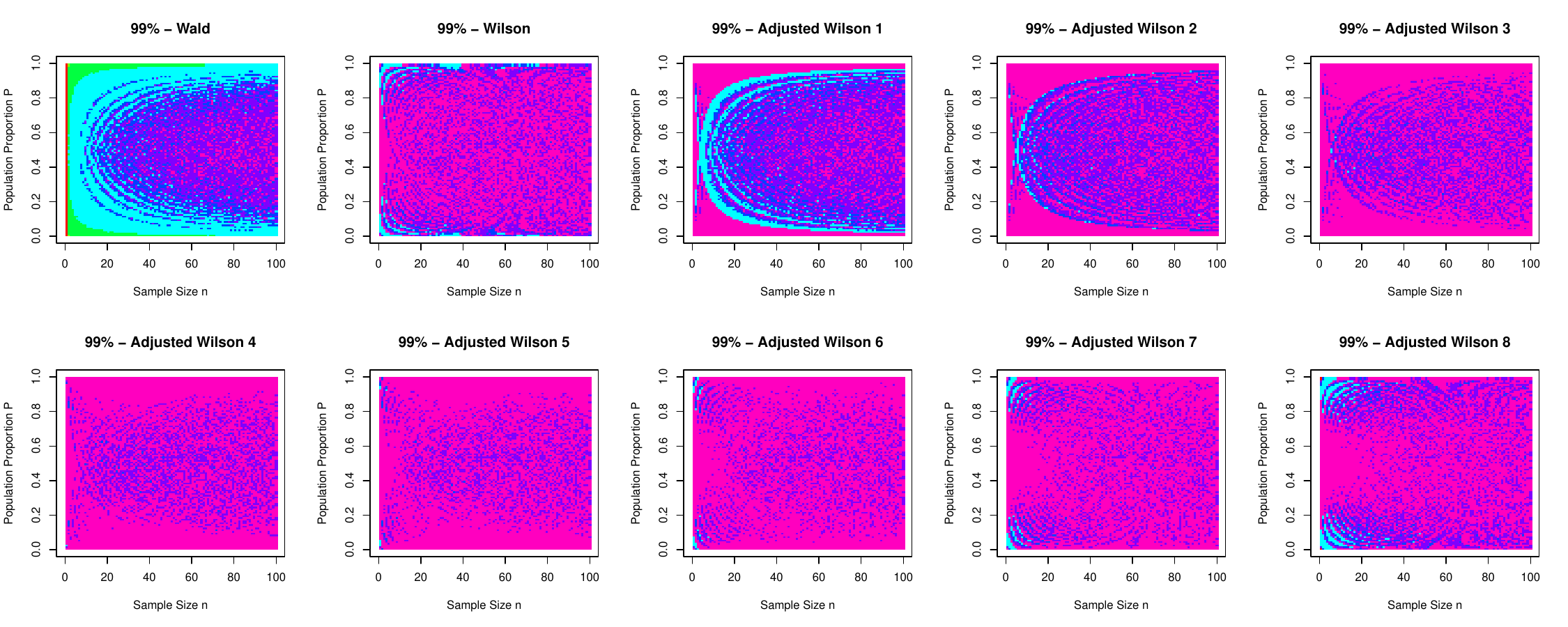}
\end{center}
\caption{Comparison of the 99\% Wald, Wilson and adjusted Wilson confidence interval's coverage probabilities for pairs $(n,p)$  where $n = 1, 2, \dots, 100$, $p = 0.01, 0.02, \dots, 0.99$}
\label{fig:99comp100}
\end{figure}

In what follows, $N_\text{Sim} = 1,000$. Let us set a counter $\kappa(n,p)$ for each of the Wald, Wilson, and adjusted Wilson of type $\epsilon$, where $\epsilon = 1, 2, \dots, 8$.  For each of the confidence levels $c = 90\%, 95\%$ and $99\%$, and for each pair (pixel) $(n, p)$, where $n=1, 2 \dots, 100$ and $p=0.01, 0.02, \dots, 0.99 $, we performed the following paired designed experiment: Generate $N = 10,000$ data points, $pN$ of which are successes and $(1-p)N$ are failures. Shuffle the data set, and randomly select a sample of size $n$. Compute the Wald, Wilson and adjusted Wilson of type $\epsilon$ confidence intervals for a population proportion, where $\epsilon = 1, 2, \dots, 8$, . For each of these ten confidence intervals, increment by one the corresponding counter $\kappa(n,p)$ whenever the population proportion $p$ belonged to the computed confidence interval. Repeat this experiment $N_\text{Sim} - 1$ more times. Finally, compute the ten relative frequencies $\pi (n,p) = \kappa(n,p)/N_{\text{Sim}}$ corresponding to the (empirical) coverage probability for a given confidence level, confidence estimator, size $n$ and population proportion $p$.  At the conclusion of this process, we obtained for each of the confidence level  $c = 90\%, 95\%$ and $99\%$, and for each of the ten confidence intervals (Wald, Wilson, and adjusted Wilson of type $\epsilon$),  $9900$ coverage probabilities $\pi(n,p)$. Figure \ref{fig:90comp100}. (resp., \ref{fig:95comp100}., and  \ref{fig:99comp100}.) contains ten pixel-colored plots where each pair (pixel) $(n,p)$ is colored according to the rainbow color  code (Table \ref{tab:ColorCode}) of its coverage probability  for a fixed confidence level $c = 90\%$ (resp., $95\%$, and $99\%$).

For each pixel-colored plot in Figures \ref{fig:90comp100}., \ref{fig:95comp100}. and \ref{fig:99comp100}., which contains $9900$ pairs (pixels) $(n,p)$, we computed the percentage of pixels that were colored in each of the eight rainbow colors. We report in Table \ref{tab:pinkcolors}. only the percentages (out of 9900) of  pairs $(n,p)$ colored in pink, that is, whose coverage probability are at least $100(1-\alpha)\%$. We define this percentage as the satisfactory pixel percentage.

\begin{table}[h]
\caption{Satisfactory pixel percentages of pairs $(n,p)$, $n = 1, 2, \dots, 100$, $p = 0.01, 0.02, \dots, 0.99$, with a coverage probability greater than or equal $100(1-\alpha)\%$.  \label{tab:pinkcolors}}
\begin{center}
\begin{tabular}{|c|c|c|c|}
\hline 
Method & 90\% CI & 95\% CI & 99\% CI\\ \hline
Wald & 0.2100 & 0.1428 & 0.0721\\
Wilson &  0.5761 & 0.6147 & 0.6027\\
Adjusted Wilson 1 & 0.4701 & 0.4127 & 0.2921 \\
Adjusted Wilson 2 & 0.6201 & 0.6022 & 0.4744\\
Adjusted Wilson 3 & \textbf{0.6926} & 0.7191 & 0.6428\\
Adjusted Wilson 4 & 0.6433 & \textbf{0.7503} & 0.7393\\
Adjusted Wilson 5 & 0.5518 & 0.6893 & 0.7866\\
Adjusted Wilson 6 & 0.4736 & 0.5800 & \textbf{0.7890} \\
Adjusted Wilson 7 & 0.4303 & 0.4879 & 0.7434 \\
Adjusted Wilson 8 & 0.3877 & 0.4382 & 0.6429\\ \hline
\end{tabular}
\end{center}

\end{table}%

For the 90\% confidence interval, the adjusted Wilson of type 3 led to the highest satisfactory pixel percentage (0.6926) of pairs $(n,p)$ whose coverage probabilities were at least $90\%$. For the 95\% confidence interval, the adjusted Wilson of type 4 led to the highest satisfactory pixel percentage (0.7503) of pairs $(n,p)$ whose coverage probabilities were at least $95\%$.  For the 99\% confidence interval, the adjusted Wilson of type 6 led to the highest satisfactory pixel percentage (0.7890) of pairs $(n,p)$ whose coverage probabilities were at least $99\%$, but the adjusted Wilson of types 5 has a very close satisfactory pixel percentage (0.7866). To determine which of the adjusted Wilson of type 5  or 6 is the best for a 99\% confidence interval for proportion, we performed the same sampling distribution comparison with 100 simulations runs. The paired t-test revealed that there is no statistical difference (p-value is 0.0818) between the satisfactory pixel percentages of pairs $(n, P)$, where $n=1, 2, \dots, 100$ and $p = 0.01, 0.02, \dots, .99$.

In light of the above paired designed comprehensive comparison, the optimal total number of pseudo-observations (half of which are successes, and the other half are failures)  for the adjusted Wilson is 3 (resp., 4 and 6) for a confidence level $c = 90\%$ (resp., 95\% and 99\%). While adding 3 (resp., 4) pseudo-observations is to be expected because the square of the critical value $z_{\alpha/2}$ is 2.7056 (resp., 3.8415) which rounds to $3$ (resp., 4) for a significance level $c = 90\%$ (resp., 95\%), the addition of 6 (or even 5, since there was no statistical difference in the satisfactory pixel percentages) pseudo-observations for the 99\% adjusted Wilson confidence level is surprising because the square of the critical value $z_{\alpha/2}$ is 6.634897 which rounds to $7$. 

\begin{figure}[h]
\begin{center}
\includegraphics[scale=.36]{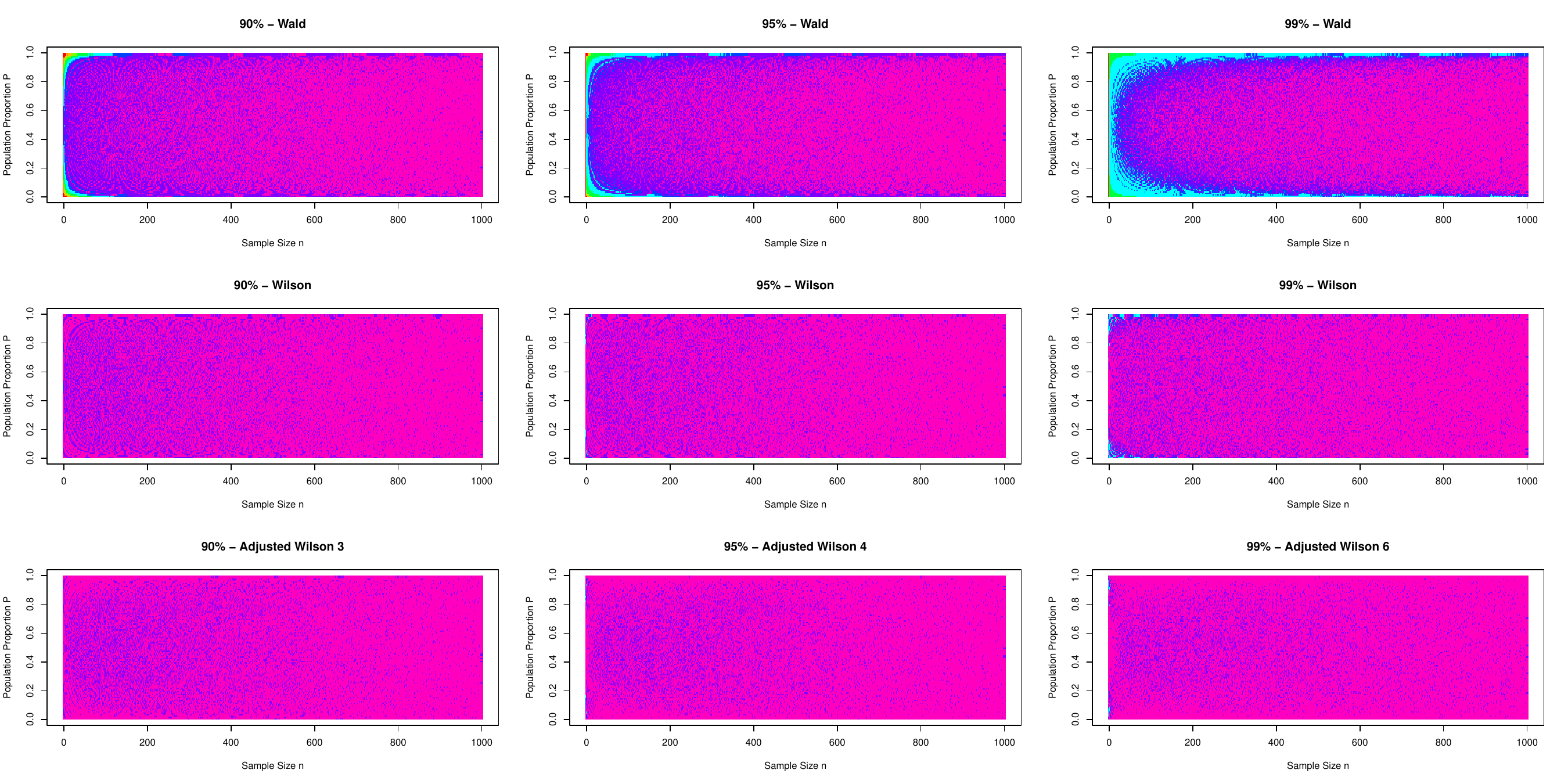}
\end{center}
\caption{Comparison of the $100(1-\alpha)\%$ Wald, Wilson and adjusted Wilson confidence interval's coverage probabilities for pairs $(n,p)$  where $n = 1, 2, \dots, 1000$ and  $p = 0.01, 0.02, \dots, .99$. \label{fig:99WWAWComp}
\label{fig:1Kallcomp}}
\end{figure}

To conclude our comparative analysis, we performed the same paired design comparison of the coverage probability for the following nine confidence intervals for a population proportion using the same number of simulations $N_{\text{Sim}} = 1,000$: the 90\%- Wald, Wilson and adjusted Wilson of type 3; the 95\% Wald, Wilson and adjusted Wilson of type 4; and the 99\% Wald, Wilson and adjusted Wilson of type 6 for sample sizes $n=1, 2, \dots, 1000$ and population proportion $p = 0.01, 0.02, \dots, 0.99$. The nine pixel-colored plots, consisting of 99,000 pixels (pairs $(n,p)$) for each plot,  are shown in Figure \ref{fig:1Kallcomp}.  Similarly, we computed the satisfactory pixel percentages, that is, the percentage (out of 99,000) of pairs (pixels) $(n,p)$ whose coverage probability are at least $100(1-\alpha)\%$. The results are summarized in Table \ref{tab:1KallSPP}.  We found that the 90\% (resp., 95\% and 99\%) adjusted Wilson of type 3 (resp., 4 and 6) confidence interval  outperformed both the 90\% (resp., 95\% and 99\%) Wald and Wilson confidence intervals. 

\begin{table}[h]
\caption{Satisfactory pixel percentages of pairs $(n,p)$,  $n = 1, 2, \dots, 100$ or $n = 1, 2, \dots, 100$, and $p = 0.01, 0.02, \dots, 0.99$, with a coverage probability greater than or equal $100(1-\alpha)\%$.  \label{tab:1KallSPP}}
\begin{center}
\begin{tabular}{|c|c|c|c|}
\hline 
Confidence Level &Method &  $n=1, \dots, 100$ & $n=1, \dots, 1000$   \\ \hline
& Wald &0.2097  & 0.6500 \\
90\% & Wilson & 0.5774&0.7708 \\
& Adjusted Wilson 3 & \textbf{0.6948}& \textbf{0.8093}\\ \hline
& Wald & 0.1391&0.6232 \\
95\% & Wilson &0.6094 &0.7969 \\
& Adjusted Wilson 4 & \textbf{0.7479}& \textbf{0.8413}\\ \hline
& Wald &  0.0723&0.5423\\
99\% & Wilson & 0.5923 &0.7819 \\
& Adjusted Wilson 6 & \textbf{0.7913} &  \textbf{0.8397}\\ \hline
\end{tabular}
\end{center}
\end{table}%

\section{Conclusion and Recommendations}
\label{sec:conc}

In light of the above analyses, summarized in Tables \ref{tab:pinkcolors} and \ref{tab:1KallSPP} and shown in Figures \ref{fig:90comp100}, \ref{fig:95comp100}, \ref{fig:99comp100} and \ref{fig:1Kallcomp}, we conclude not only that adding 3 (resp., 4 and 6) pseudo-observations is the best for the 90\% (resp., 95\% and 99\%) adjusted Wilson confidence interval, but it also performs better than both the 90\%  (resp., 95\% and 99\%) Wald and Wilson confidence intervals.  

This comprehensive comparison of the Wald, Wilson and adjusted Wilson confidence interval expands and completes the work of \cite{Agresti1} in which they compared the mean coverage performance, that is averaging the coverage probability across all population proportion in $(0,1)$, for sample sizes $n=5, 15, 30, 50$ and $100$ and for only the 95\% confidence level, to not only all sample sizes $n=1, 2, \dots 100$ (and $1000)$ and population proportion $p=0.01, 0.02, \dots, 0.99$, but also for the three typical confidence levels 90, 95 and 99\%.

The author encountered the adjusted Wilson confidence interval for the first time in \cite{SLS}, a textbook that author used for teaching an applied and computational statistics course tailored for life science students. Since the publication of \cite{Agresti1} and \cite{Agresti2}, it is surprising and quite unfortunate that the number of introductory and intermediate statistics textbook introducing the Wilson and/or the adjusted Wilson confidence interval is a small single digit, which includes \cite{GTSIBP},  \cite{MSwRaR} and  \cite{SLS}. We do hope that this comprehensive analysis will help convince educators to introduce the adjusted Wilson (of types 3, 4 and 6 for the 90, 95, and 99\% confidence level respectively) confidence intervals in their statistics classes, and for statistics practitioners to systematically use it in their analysis, especially for small sample sizes and/or when the population proportion is close to 0 or 1.











\bibliographystyle{Chicago}

\bibliography{WWAWBiblio}
\end{document}